# High Energy Colliding Beams; What Is Their Future?[1]

Burton Richter

Stanford University and SLAC National Accelerator Laboratory

brichter@slac.stanford.edu

## Abstract

The success of the first few years of LHC operations at CERN, and the expectation of more to come as the LHC's performance improves, are already leading to discussions of what should be next for both proton-proton and electron-positron colliders. In this discussion I see too much theoretical desperation caused by the so far unsuccessful hunt for what is beyond the Standard Model, and too little of the necessary interaction of the accelerator, experimenter, and theory communities necessary for a scientific and engineering success. Here, I give my impressions of the problem, its possible solution, and what is needed to have both a scientifically productive and financially viable future.

1: Introduction

I have been asked to introduce this issue on what might be beyond the frontier colliding-beam machines of today; the proton-proton Large Hadron Collider (LHC) currently operating at CERN; and the electron-positron International Linear Collider (ILC), which though fully designed, is still only a dream in the minds of its proponents.

I think I may be the last still around of the first generation of pioneers that brought colliding beam machines to reality. I have been personally involved in building and using such machines since 1957 when I became part of the very small group that started to build the first of the colliders[2]. While the decisions on what to do next belong to the younger generation, the perspective of one of the old guys might be useful. I see too little effort going into long range accelerator R&D, and too little interaction of the three communities needed to choose the next step, the theorists, the experimenters, and the accelerator people. Without some transformational developments to reduce the cost of the machines of the future, there is a danger that we will price ourselves out of the market.

---

[1] To be published in *Review of Accel. Sci. & Technology*, Vol. 7 (2014), Editors Alexander W. Chao and Weiren Chou,, World Scientific Pub., Singapore.
[2] If you are interested in the history of the development of the colliders from the beginning to today, in 1992 I wrote a long article titled "The Rise of Colliding Beams." It can be found in "The Rise of the Standard Model", Lillian Hoddeson et al., Cambridge University Press, 1997, and as a SLAC publication SLAC-PUB-6023. It has detail on who really did what.

*Work supported in part by US Department of Energy under contract DE-AC02-76SF00515.*

**SLAC National Accelerator Laboratory, Stanford University, Stanford, California 94039, USA**

In section 2 below I will give a short review of how we have come to where we are in the 58 years since the first discussion of the potential of colliding beam machines. The development of that technology is what has allowed the field to reach the energies of today's frontier facilities.

In section 3 I go on to discuss the next step in proton colliders and find that there seems to be a great danger of setting the luminosity of the 100 TeV example being discussed today too low and so severely limiting its discovery potential.

Section 4 takes a look at the future of electron colliders. The linear collider technology being proposed for the next facility seems to me to allow for the construction of the first and last machine of this type because, while the cost of the first seems affordable, the cost of the next which has to have much higher energy will not be using current technology.

Section 5 gives some personal thoughts on what needs to be done if accelerator based particle physics is to have a long term future.

Section 6 concludes the paper with some personal thoughts on theory, experiment, science politics, and problems looming for large international collaborations.

2: A Short Look Back

The beginning of colliders came in a paper by D.W. Kerst and his Midwest University Research Association (MURA) group, published in 1956[3] followed by a longer paper in the Proceedings of the CERN Accelerator Conference[4], also in 1956. At the time the highest energy proton accelerators were the 3.3-GeV Cosmotron at Brookhaven National Laboratory, and the 6.2-GeV Bevatron at the Lawrence Berkeley National Laboratory. It seemed as if new baryon and meson states were being discovered almost as fast as data could be collected and analyzed, and both Brookhaven and CERN were discussing building accelerators with an energy of about 25 GeV. Kerst wrote that it might be possible to go far beyond those energies with a new accelerator technology, pointing out that making the beams from two 21.6-GeV accelerators collide head-on would give a center-of-mass energy equivalent to that of one accelerator of 1000 GeV colliding with a proton fixed target.

---

[3] D.W. Kerst et al., Phys. Rev. 102, 590 (1956)
[4] CERN Symposium on High Energy Accelerators and Pion Physics (Geneva: CERN, 1956), p.36.

At this same symposium a new actor came on stage, G. K. O'Neill of Princeton University. He too was interested in proton-proton collisions at very high center-of-mass energies, and he introduced the notion of the accelerator-storage ring complex. Beams would be accelerated to some high energy in a synchrotron and then transferred into two storage rings with a common straight section where the beams would interact. Since the beams at high energy need much less space in an accelerator vacuum chamber than is required for beams at injection, the high-energy storage rings would have smaller-cross-section magnets and vacuum chambers, thus adding little to the cost of the complex, but at the same time enormously increasing the scientific potential. He also observed, "The use of storage rings on electron synchrotrons in the GeV range would allow the measurement of the electron-electron interaction at center-of-mass energies of about 100 times as great as are now available. The natural beam damping in such machines might make beam capture somewhat easier than in the case of protons." That observation had a profound effect on O'Neill's career and mine, as well as on particle physics.

How to realize a colliding-beam machine was the question. The MURA FFAG accelerators discussed by Kerst were enormously complex, and none had ever been built at that time (only one small one has been built since). There was considerable concern about whether FFAG machines would actually work as well as their proponents claimed.

At the same time, the problem of injection into the proton-synchrotron storage-ring complex that O'Neill and others discussed was thought to be very difficult. Indeed, O'Neill's original idea of using a scattering foil for injection was soon proved to be impossible. On the other hand, injection and beam stacking in an electron storage ring looked easy because of synchrotron-radiation damping. An electron beam could be injected off-axis into a storage ring and would perform betatron oscillations around the equilibrium orbit. These oscillations would decrease exponentially over time in a properly designed magnet lattice because of the emission of synchrotron radiation while the energy loss is compensated with RF acceleration. When these oscillations had damped sufficiently, another bunch could be injected into the storage ring and would damp down on top of the first one. Since phase-space was not conserved in the presence of radiation, there was, in principle, no stacking problem. Thus, the collider story began with electron machines because they were much easier to inject into, were smaller, and were much less expensive than the proton machines that were the dream of those who first began the move toward colliders.

In 1957 O'Neill visited Stanford's High Energy Physics Lab (HEPL) to discuss the possibility of building an electron-electron collider with W.K.H. Panofsky, its Director. O'Neill wanted to test the technology and to test Quantum Electrodynamics (QED) to a level far beyond anything that had been done before. Panofsky was very interested and began helping to put together a Princeton-Stanford collaboration to make a real proposal for such a facility, which is where I came in. I was a post-doc interested in testing quantum electrodynamics and happily joined the adventure to bring to life a new tool for high-energy physics research, and to allow a test of QED to a point far beyond what had been done earlier by me.

The group became O'Neill; Bernard Gittelman, a Princeton post doc; W.C. (Carl) Barber, a Stanford senior staff member; and me. Panofsky convinced the Office of Naval Research to fund the project and in December of 1958 the construction of the first colliding beam machine (CBX) began, launching the collider era. The first beam was stored on March 28, 1962; the first physics results testing QED were presented in 1963; and the facility was finally shut down in 1968.

It took longer for the proton colliders to begin. The information on the beam-beam interaction from the electron collider was important, but even more important was a great deal of theoretical and experimental work on how to inject and stack beams in a proton ring. CERN began what was to be the ISR in 1966 and first collisions began in 1971. The colliding-beam era was fully launched.

In the nearly 60 years since the first suggestions on the potential of colliding beams were made, enormous progress has occurred in both accelerator center-of-mass energies (to 200 GeV in the e+e- system and 8 TeV in the p-p system), and in our understanding of matter, energy and the structure of our universe. Colliders have played a major role, but fixed-target machines have been important too (neutrino physics, for example), as have ground- and space-based instruments (dark matter and dark energy, for example).

We have our Standard Model which explains a lot, but not all. It is like a beautiful manuscript with Post-it notes stuck on pages here and there. CP violation is allowed, but not required; neutrino masses and oscillations are allowed but not explained; dark matter is allowed, but what it is is a mystery; the particle-antiparticle asymmetry of the universe is allowed but not explained; etc. Taking down some of those notes and gaining a deeper understanding of our universe is the object of higher-energy colliding-beam systems.

In Asia, Europe and the U.S. scientists and funding agencies have been trying to set priorities on next steps for what has come to be called the energy, intensity, and cosmic frontiers. Progress on all is needed to deepen our understanding. This journal issue is about proceeding on the energy frontier, but it is important not to forget that with what we know today, there are other areas of importance as well. The energy frontier will not, and should not, get all the money.

3: A Look Ahead to Beyond LHC-2014

In the early 1980s the U.S. was beginning to develop the ideas for what became known as the Superconducting Super Collider (SSC), a 40-TeV center-of-mass (CM) superconducting p-p colliding-beam machine. The first mention I know of what became the LHC is in an internal report from the LEP group (LEP note 440, 1983) by Stephen Myers and Wolfgang Schnell about putting a superconducting p-p collider in the LEP tunnel after the LEP e+e- collider had done its job. The LEP tunnel was only roughly one-third the circumference of the SSC tunnel, but that could be partly made up for with higher field superconducting magnets (eventually the LHC magnets ran at about 30% higher field than the design field of the SSC).

The Myers/Schnell paper started informal discussions at CERN that became more serious when the SSC was initially approved by the U.S. Congress, and turned into a major design effort when the SSC was canceled by the U.S. Congress in 1993.[5] Construction of the LHC began in 1998 and first operational trials began in 2008. Unfortunately, a magnet short and massive quench of the superconducting systems damaged the machine and the real start of operations at about only half the design energy began in early 2010. In early 2015 the LHC will begin operations again at about 13 TeV compared to the 8-TeV operations before its recent shutdown for upgrading.

The LHC itself is an evolving machine. Its energy at its restart next year will be 13 TeV, slowly creeping up to its design energy of 14 TeV. It will shut down in 2018 for some upgrades to detectors, and shut down again in 2022 to increase the luminosity. It is this high-luminosity version (HL-LHC) that has to be compared to the potential of new facilities. There has been some talk of doubling the energy of the LHC (HE-LHC) by replacing the 8-tesla magnets of the present machine with 16-tesla magnets, which would be relatively easy compared to the even more talked about bolder step to 100 TeV for the next project. It is not clear to me why 30-TeV LHC excites so little interest, but that is the case.

A large fraction of the 100 TeV talk (wishes?) comes from the theoretical community which is disappointed at only finding the Higgs boson at LHC and is looking for something that will be real evidence for what is actually beyond the standard model. Regrettably, there has been little talk so far among the three communities, experimenters, theorists, and accelerator scientists, on what constraints on the next generation are imposed by the requirement that the experiments actually produce analyzable data. I will give an example of an LHC-like 100-TeV machine (LHC-100) to show some of the design problems.

---

[5]If you want to know more about the rise and fall of the SSC see Michael Riordan, Lillian Hoddeson and Adrienne W. Kolb, *Tunnel Visions: The Rise and Fall of the Superconducting Super Collider*, (University of Chicago Press, to be published in 2015)

The most important choice for a new, higher energy collider is its luminosity, which determines its discovery potential. If a new facility is to have the same potential for discovery of any kind of new particles as had the old one, the new luminosity required is very roughly proportional to the square of the energy because cross sections typically drop as $E^{-2}$. A seven-fold increase in energy from that of HL-LHC to a 100-TeV collider therefore requires a fifty-fold increase in luminosity. If the luminosity is not increased, save money by building a lower-energy machine where the discovery potential matches the luminosity.

All the studies that I have seen so far on the 100TeV collider have much lower luminosities, more like that of HL-LHC, and limit themselves to events per beam crossing about the same as that machine will have. Even at only the luminosity of HL-LHC there are some new things that can be found at the new very high-energy facility. Examples are new gauge bosons like a Z', or Gluino where particles with masses up to 30-TeV for the weak boson, or 10-TeV – 15-TeV for the Gluino could be found (if any were there). Of course, restricting the luminosity to what will be achieved at HL-LHC gives the new machine a limited vision, and will (and should) seriously lower the likelihood that it will be funded[6].

Table 1 below gives key parameters for my examples scaling up to both high luminosity and energy from HL-LHC (24 July, 2014 parameter list) with two versions of the LHC-100, one with 8-tesla magnets and the other with 16-tesla magnets (16 tesla can be gotten with niobium-tin conductor or with high-Tc magnets and that choice has big implications for the cryogenic system that I will not go into here). The parameters for the high-energy machines are workable, but not optimized. An optimized machine will surely be different, and that optimization will be needed to make the machine more affordable and the experiments more likely to succeed.

All other parameters of importance to the accelerator and experimenter communities are linked to the luminosity number. I have kept the bunch spacing of the LHC. Double the spacing and the number of particles per bunch has to go up, as does the number of interactions per beam crossing, while the beam current and the synchrotron radiation power goes down. Halve the bunch spacing and the opposite occurs; particles per bunch and interactions per crossing go down while beam current and synchrotron radiation go up.

I use the same value of β* as in HL-LHC, though it may be difficult to make it that small at the much higher energy. I also assume crab crossing is used. Also, using an injector chain similar to the LHC's, seven times as many bunches have to be stacked, though adiabatic damping going to higher energy matches the increase in phase space giving the same size beam.

---

[6] There are some advocating scaling the luminosity of a 100-TeV machine from the design luminosity of the SSC. I would remind those so advocating, that the SSC was not built, the LHC was.

Table 1: Examples of 100 TeV colliders scaled from HL-LHC

| Parameter | HL-LHC | LHC-100 8T | LHC-100 16T |
|---|---|---|---|
| Beam Energy (TeV) | 7 | 100 | 100 |
| Circumference (km) | 27 | 190 | 95 |
| L (cm$^{-2}$sec$^{-1}$) | 5x10$^{34}$ | 2.5x10$^{36}$ | 2.5x10$^{36}$ |
| Bunch Spacing (ns) | 25 | 25 | 25 |
| Beam Current (Amp) | 1.09 | 7.7 | 7.7 |
| Synchrotron Rad Power (Mw) | 0.0075 | 2.6 | 10.3 |
| β * (cm) | 15 | 15 | 15 |
| ε$_n$ (micron) | 2.5 | 2.5 | 2.5 |
| Particles per Bunch | 2.2x10$^{11}$ | 1.5x10$^{12}$ | 1.5x10$^{12}$ |
| Events per bunch collision | 140 | 7000 | 7000 |
| Events per mm | 1.3 | 0.0025 | 0.0025 |

The events per beam crossing and per unit length along the collision region are going to make serious problems for the detectors. Having 50 times the events per beam crossing will require something new in detectors. Having the mean spacing between vertices go from 1.3 mm to 2.5 microns will probably also require something new in detector technology. Getting the experimenters involved in setting parameters is necessary in building something that can really do the physics[7].

I understand that CERN is setting up such a group. It is about time someone did so, and a serious discussion on its physics reach should be the first order of business. Detector R&D will be as important in the long run as machine R&D.

4: A Look Ahead to the ILC and Beyond

Though the first storage ring electron collider started operations about ten years before the first proton collider, the proton machines have gone far beyond their electron cousins in energy. The reason is the same one that made electron colliders easier to build initially, synchrotron radiation. While synchrotron radiation made it easy to inject into an electron collider, it made it much more expensive to go to very high energy. The problem was that the minimum-cost machine increases in circumference and cost as the square of the energy.[8] I used to use a slide based on a scale-up of the 100-GeV

---
[7] The summary talk by Weiren Chou at the Fermilab Hadron Collider Workshop, August 25-28, 2014, is an excellent introduction to the issues; see
https://indico.fnal.gov/conferenceOtherViews.py?view=standard&confId=7864
[8] Burton Richter, *Very High Energy Electron-Positron Colliding Beams for the Study of Weak Interactions*, Nuclear instruments and Methods, 136 (I976) 47-60

version of LEP to 1-TeV energy that I called LEP 1,000.  It had a 2700-km circumference with one interaction region in Geneva and the other in London.

I started thinking about an alternative, and in 1978 at an accelerator conference at Fermilab I found that A. N. Skrinsky (Novosibirsk), M. Tigner (Cornell), and I had been thinking about the same issue.[9]  We worked out the limitations (primarily from the very strong beam-beam interaction that came to be called beamstrahlung), and when I returned to SLAC began the design of a system to test the concept that used only one existing linac and so would be affordable.  Both electron and positron bunches would be accelerated in the same pulse of the SLAC linac, and brought them into collision using a system of magnets.  The entire thing was shaped something like a tennis racquet in which both electron and positron bunches would be accelerated in the same linac pulse, split at the end of the linac and send each on its way to the collision point around the head of the racquet.  An imaginative Department of Energy invented a budget category they called an R&D construction project to fund it.  First collisions were in 1989 and by 1992 with longitudinally polarized beams the SLAC Linear Collider (SLC) was producing real physics, and the accelerator physics community began to think about big linear colliders.

In 1993 the SSC project was canceled by the U.S. Congress.  One of the reasons was that no other country had agreed to contribute to the construction of the project.  Hirokata Sugawara (Director of KEK), Bjorn Wiik (Director of DESY), and I (Director of SLAC) discussed how to move linear colliders along in the light of the crash of the SSC and concluded that one of the SSC's problems was that potential collaborators were not part of the group that decided on the parameters.  We thought we might do better if we worked together rather that separately and thereby came up with a machine design that had international backing from the very beginning.  It would not guarantee collaborations, but would eliminate one of the barriers to such that affected the SSC.

This worked for a while, but broke down when it came to deciding on the technology for the design study.  We had let technologies become associated with laboratories so choosing the better option between superconducting and room-temperature RF was not as easy as it would have been if each of the three labs had been involved with both of the technologies.  The same mistake is being made again: CERN works on CLIC while the other labs work on superconducting linac systems.

An international group was created to design and estimate the cost of a superconducting linear collider with initial center-of-mass energy of 500 GeV,

---

[9] L. E. Augustin et al, *Limitation of Performance of e+e- Storage Rings and Linear Colliders Systems at High energy,* Proceedings of the ICFA Workshop on Possibilities and Limitations of Accelerators and Detectors, Fermi National Accelerator Laboratory, 1978

expandable to 1 TeV at a later date.  The Reference Design Report with cost estimate was delivered in 2007.  In 2007 U.S. dollars the cost was estimated to be $6.6 billion including site specific costs, plus 24 million person hours of labor.  Accounting was of the kind used in Europe and Japan where only the costs of contracts is normally included.  U.S. practice includes much more such as R&D before and during construction, detectors, escalation, contingency, and all labor including that at the laboratories where the work was to be done and managed.  Typically this increases costs over the accounting method used at the ILC by a factor of 2.5 to 3, resulting in a cost estimate of $16 to $20 billion for the project in U.S. terms.  This was about the inflation-adjusted cost estimate for the SSC when it was canceled in 1993.  Further, in 2008 the financial crash made it even more unlikely that a project that large would be funded even as an international effort.  Work on the ILC slowed dramatically.

The discovery of the Higgs boson at the LHC in 2012 raised interest in a somewhat lower energy e+e- collider to be used as a kind of Higgs factory to produce lots of them through the Z-Higgs channel.  The scientific advances would come from precision measurement of Higgs branching fractions.  The standard model Higgs couples to leptons proportional to the square of their masses.  Deviations from the standard predictions give clues to what might be waiting to be discovered through internal loop diagrams if masses are above what can be seen in Higgs decays, or even directly if the mass of something new is low enough (not likely since if so, it should have been found at the LHC).

The precision being discussed is at about the 1% level for each of the Higgs branching fractions being investigated.  This is very difficult because the Z-Higgs channel has to be tagged by finding the Z decay and then finding the tau pair decays of the higgs, and may be impossible for the more complex bottom and charm quark decays of the higgs.  The problem will be backgrounds as well as statistics.  If the theorists really need 1% branching fraction, do not go ahead until the experimenters say if it can be done to that precision in the real world of many open channels and only a relatively small Higgs production.  Also, check the luminosity needed.

5: The Long Term Future of Accelerator Based Physics

When I was much younger I was a fan of science fiction books.  I have never forgotten the start of one, though I don't remember the name of the book or its author.  It began by saying that high-energy physics' and optical astronomy's instruments had gotten so expensive that the fields were no longer funded.  That is something that we need to think about.  Once before we were confronted with a cost curve that said we could never afford to go to very high energy, and colliding beams were invented and saved us from the fate given in my science fiction book.  We really need to worry about that once more.

In the proton collider world, building the LHC, its detectors, and the repairs required to get it to its design energy will cost something like $10 billion. The next step being discussed entirely too casually by some of the theorists would take us to 100 TeV (CM) at what should be a luminosity 50 times that of the still non-existing HL-LHC. If the cost of the next-generation proton collider is really linear with energy, I doubt that a 100-TeV machine will ever be funded, and the science fiction story of my youth will be the real story of our field.

I see no well-focused R&D program looking to make the next generation of proton colliders more cost effective. I do not understand why there is as yet no program underway to try to develop lower cost, high-Tc superconducting magnets done on the scale of R.R. Wilson's efforts at Fermilab to successfully develop the first generation of commercially viable superconducting magnet that led to the Tevatron, Hera and LHC. The only place that might do it today is CERN. I hope they try.

I am both more optimistic and more pessimistic about e+e- colliders. More optimistic because accelerating gradients of more than 50 GeV per meter (50 TeV per kilometer sounds even more exciting) have already been demonstrated in plasma-wakefield acceleration and of several GeV per meter in laser acceleration, though both have now poor 6-dimensional phase space; more pessimistic because I don't see a push to develop these technologies for use in real machines.

The e+e- colliders have two advantages over the proton colliders. The cross sections of interest are all of comparable orders of magnitude. The background of 10 billion or more uninteresting events for each interesting one, the problem of proton colliders, does not really exist for the electron colliders. There is a low transverse momentum fizz that is confined to small angle, but the interesting events are much easier to get at. In addition the equivalent mass reach in the electron colliders only requires 10% to 20% of the energy of the proton collider with the same mass reach. The 100-TeV p-p collider is matched by a 10- to 20-TeV electron collider.

My challenge to the electron accelerator community is to produce a cost effective system with an acceleration gradient of at least 1-GeV per meter with reasonable transverse phase space and an energy spread of no more than 10% to 20%. Because of the parton distribution in the proton, the effective energy spread in p-p collisions is more like 100%. You have about 15 years to do it since that is the time to when HL-LHC will start to operate.

## 6: A Few Final Thoughts

The usual back and forth between theory and experiment; sometimes one leading, sometimes the other leading; has stalled. The experiments and theory of the 1960s and 1970s gave us today's Standard Model that I characterized earlier as a beautiful manuscript with some unfortunate Post-it notes stuck here and there with unanswered questions written on them. The last 40 years of effort has not removed even one of those Post-it notes. The accelerator builders and the experimenters have built ever bigger machines and detectors, while the theorists have kept inventing extensions to the model.

If you have seen the movie *Particle Fever* about the discovery of the Higgs boson, you have heard the theorists saying that the only choices today are between Super-symmetry and the Landscape. Don't believe them. Super-symmetry says that every fermion has a boson partner and vice versa. That potentially introduces a huge number of new arbitrary constants which does not seem like much progress to me. However, in its simpler variants the number of new constants is small and a problem at high energy is solved. But, experiments at the LHC already seem to have ruled out the simplest variants.

The Landscape surrenders to perpetual ignorance. It says that our universe is only one of a near infinity of disconnected universes, each with its own random collection of force strengths and constants, and we can never observe or communicate with the others. We can never go further in understanding because there is no natural law that relates the different universes. The old dream of deriving everything from one constant and one equation is dead. There are two problems with the landscape idea. The first is a logic one. You cannot prove a negative, so you cannot say that there is no more to learn. The second is practical. If it is all random there is no point in funding theorists, experimenters, or accelerator builders. We don't have to wait until we are priced out of the market, there is no reason to go on.

There is a problem here that is new, caused by the ever-increasing mathematical complexity of today's theory. When I received my PhD in the 1950s it was possible for an experimenter to know enough theory to do her/his own calculations and to understand much of what the theorists were doing, thereby being able to choose what was most important to work on. Today it is nearly impossible for an experimenter to do what many of yesterday's experimenters could do, build apparatus while doing their own calculations on the significance of what they were working on. Nonetheless, it is necessary for experimenters and accelerator physicists to have some understanding of where theory is, and where it is going. Not to do so makes most of us nothing but technicians for the theorists. Perhaps only the theory phenomenologists should be allowed to publish in general readership journals or to comment in movies.

The ever rising cost of the ITER fusion-energy project has increased skepticism that any big international project can be brought in on budget. I have had to deal with more than one official skeptic in Washington on the potential of large-scale international collaborations. I have pointed to the LHC as a counter-example. ITER was to be built with the nations involved contributing components that each country committed to build. Its central management had no money and no authority, so could not bring additional resources to bear when one important part of the project fell behind, and could do nothing to move components from one country to another to solve production problems. The result has been a more than tripling of the ITER cost estimate and a delay of more than a decade in completion.

At LHC almost all the funds were managed centrally with the understanding (never formally stated) that each country that contributed to the project would see contracts roughly in proportion to its contribution. The LHC model worked while the ITER model has not. When we think of a next very large international accelerator project we have to think of a management system that will not result in an ITER like financial and scheduling disaster.

I end with best wishes to the younger generation. May you make real progress to the one constant, one equation solution to the question that brought me to HEP: how does the universe work?